\documentclass[prl,twocolumn,showpacs,amsmath,amssymb]{revtex4}

\usepackage{graphicx}
\usepackage{subfigure} 

\begin{document}

\title{Classical to Quantum Transition of a Driven Nonlinear 
  Nanomechanical Resonator} 

\author{Itamar Katz}%
\author{Alex Retzker}%
\altaffiliation[Current address:\ ]{Institute for Mathematical
  Sciences, Imperial College London, SW7 2PE, UK.}
\author{Raphael Straub}%
\altaffiliation[Permanent address:\ ]{Department of Physics,
  University of Konstanz, D-78457 Konstanz, Germany.}
\author{Ron Lifshitz}%
\altaffiliation[Currently on leave at:\ ]{Condensed Matter Physics
  114-36, California Institute of Technology, Pasadena CA 91125, USA}
\email{ronlif@tau.ac.il}
\affiliation{School of Physics \& Astronomy, Raymond and Beverly
  Sackler Faculty of Exact Sciences, Tel Aviv University, 69978 Tel
  Aviv, Israel}

\date{February 8, 2007}

\begin{abstract}
  We seek the first indications that a nanoelectromechanical system
  (NEMS) is entering the quantum domain as its mass and temperature
  are decreased. We find them by studying the transition from
  classical to quantum behavior of a driven nonlinear Duffing
  resonator.  Numerical solutions of the equations of motion,
  operating in the bistable regime of the resonator, demonstrate that
  the quantum Wigner function gradually deviates from the
  corresponding classical phase-space probability density. These clear
  differences that develop due to nonlinearity can serve as
  experimental evidence, in the near future, that NEMS resonators are
  entering the quantum domain.
\end{abstract}

\pacs{03.65.Ta, 85.85.+j, 05.45.-a, 03.65.Yz}
\maketitle%

The race to observe quantum-mechanical behavior in human-made
nanoelectromechanical systems (NEMS) is bringing us closer than ever
to testing the basic principles of quantum
mechanics~\cite{blencowe04,schwab_roukes05}. With recent experiments
coming within just an order of magnitude from the ability to observe
quantum zero-point motion~\cite{knobel03,LaHaye04,Naik06}, ideas about
the quantum-to-classical
transition~\cite{Penrose96,Leggett99,Leggett02} may soon become
experimentally accessible, more than 70 years after Schr\"{o}dinger
described his famous cat paradox~\cite{cat}. As nanomechanical
resonators become smaller, their masses decrease and natural
frequencies $\Omega$ increase---exceeding 1GHz in recent
experiments~\cite{HZMR03,cleland04}. For such frequencies it is
sufficient to cool down to temperatures on the order of $50mK$ for the
quantum energy $\hbar\Omega$ to be comparable to the thermal energy
$k_{B}T$.  Cooling down to such temperatures should allow one to
observe truly quantum mechanical phenomena, such as resonances,
oscillator number states, superpositions, and
entanglement~\cite{peano04,Carr01,Armour02,santamore04,Eisert04}, at
least for macroscopic objects that are sufficiently isolated from
their environment.


In this Letter we seek the first experimental signatures, indicating
that a mechanical object is entering the quantum domain as its mass
and temperature are decreased. Are there any clear quantum-mechanical
corrections to classical dynamics that we can hope to observe even
before the fully-quantum regime is accessible experimentally?  To
answer this question we perform two separate calculations on the same
physical object---a driven nanomechanical resonator---viewing it once
as a quantum-mechanical system and once as a classical system.  With
everything else being equal, this allows us to contrast the dynamics
of a classical resonator with that of its quantum clone. We can then
search for a regime in which quantum dynamics just begins to deviate
away from classical dynamics, providing us with the experimental
evidence we are looking for.

We calculate the dynamics numerically. We start each quantum
calculation with a coherent state, which is a minimal wave-packet
centered about some point in phase space. We start the corresponding
classical calculation with an ensemble of initial
conditions---typically of $N=10^{4}$ points---drawn from a Gaussian
distribution in phase space that is identical to the initial
quantum-mechanical probability density. We display the calculated
quantum dynamics in phase space using the quantum Wigner function
\begin{equation}\label{wigner}
W(x,p,t) = {1\over{\pi\hbar}} \int_\infty^\infty dx'
e^{-{2i\over\hbar}px'} \langle x+x'|\rho(t)|x-x'\rangle,
\end{equation}
where $\rho(t)$ is the density operator, and compare it with the
time-evolution of the classical phase-space density. We remind the
reader that the Wigner function is not a true probability distribution
as it may possess negative values, particularly when the quantum state
has no classical analog. Nevertheless, its governing equation of
motion reduces to the classical Liouville equation upon formally
setting $\hbar=0$, or whenever the potential is no more than
quadratic. Moreover, it reduces to the quantum probability $P(x,t)$ of
observing the system at position $x$ at time $t$ upon integration over
$p$, and {\it vice versa.}

We perform our calculations for three qualitatively different
situations: (1) An isolated resonator with no coupling to the
environment; (2) A resonator coupled to a heat bath at temperature
$T_{env}=0$; and (3) A resonator coupled to a heat bath at a finite
temperature $T_{env}>0$. In all cases we expect to find a regime in
which the evolution of quantum observables agrees with the
corresponding classical averages at least up to the so-called
``Ehrenfest time''~\cite{berry78,cametti02}, although it is generally
difficult to strictly define this time scale~\cite{oliveira03,%
  oliveira06}. For the isolated resonator, we do not expect to see a
convergence of the two phase-space distributions, as the limit
of $\hbar\rightarrow0$ is singular~\cite{berry89,habibPRL}.  When we
couple the resonator to an environment, as in a real experiment, we do
expect to see a gradual classical to quantum
transition~\cite{habibPRL,greenbaum05,habibLANL}.  We wish to study
the details of this transition here.

As mentioned above, the quantum dynamics of a coherent state in a
harmonic potential is essentially classical. We therefore consider a
driven Duffing resonator---a nonlinear resonator commonly observed in
experiments~\cite{Craighead00,Inna05,Inna06,aldridge05}, whose
Hamiltonian is given by
\begin{equation}\label{Hamiltonian}
H_{sys}=\frac{1}{2}p^{2} + \frac{1}{2}x^{2} -
xF\cos\omega t + \frac{1}{4}\varepsilon x^{4},
\end{equation}
where the mass $m$ of the resonator, and its natural frequency
$\Omega$ have been scaled to 1 by an appropriate choice of units.
Thus, we measure the degree in which we approach the quantum domain by
an increasing effective value of $\hbar$, as measured in the scaled
units. As we shall see, quantum corrections clearly appear for values
of $\hbar\simeq 0.1$. We shall re-interpret these values later in
terms of real experimental masses and frequencies.  The remaining
parameters in the Hamiltonian are the driving amplitude $F$, the
driving frequency $\omega$, and the non-linearity parameter
$\varepsilon$.  For the calculations presented here we take
$\varepsilon=0.01$, $F=0.015-0.06$, and $\omega=1.016-1.02$.  For this
choice of parameters the classical Duffing resonator is in the
bistability regime, where in the presence of dissipation it can
oscillate at one of two different amplitudes, depending on its initial
conditions.

\begin{figure}[bt]
    \begin{center}
    \subfigure[\ Initial coherent state, $t=0$]{
    \includegraphics[width=0.23\textwidth]{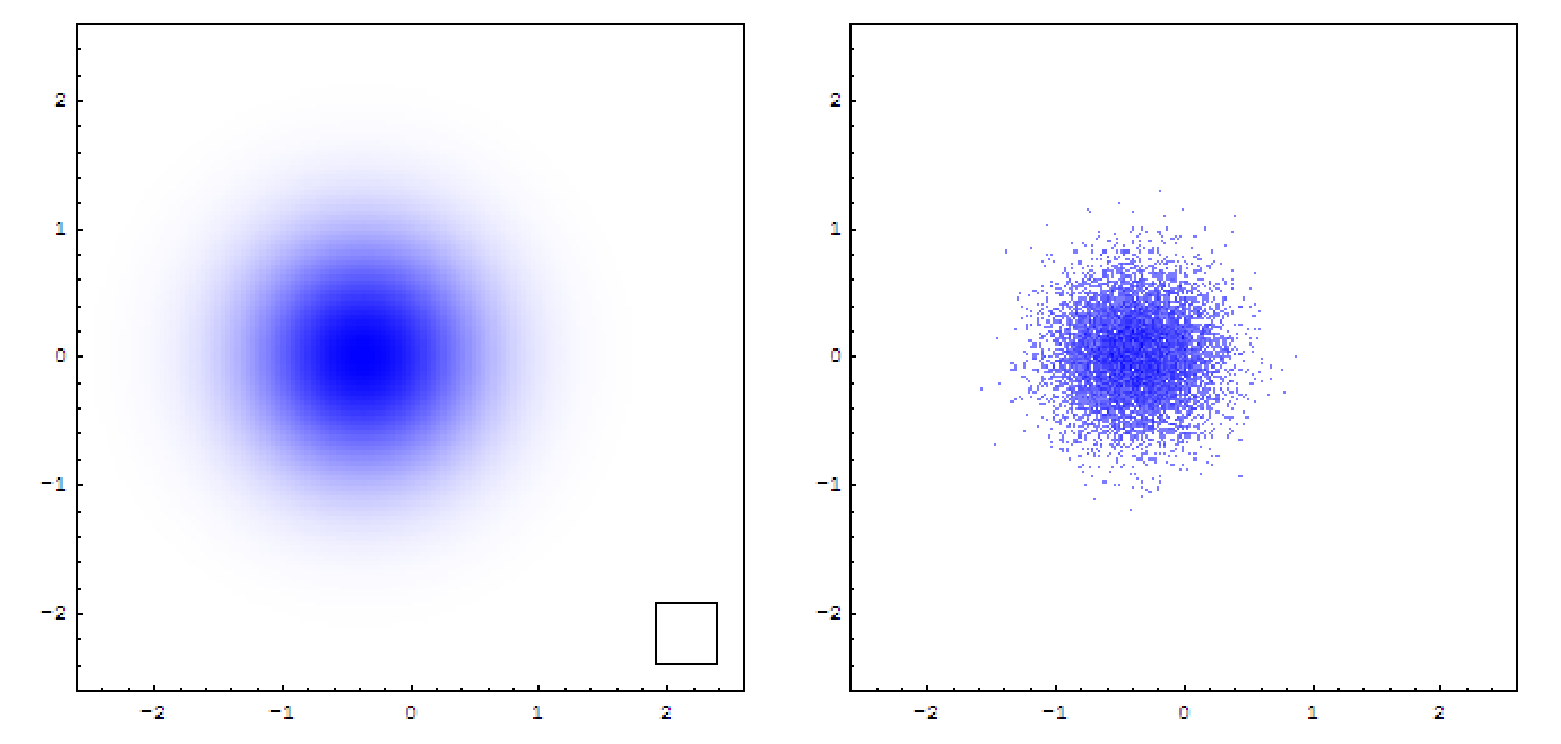}}
    \subfigure[\ $t=30T$]{
    \includegraphics[width=0.23\textwidth]{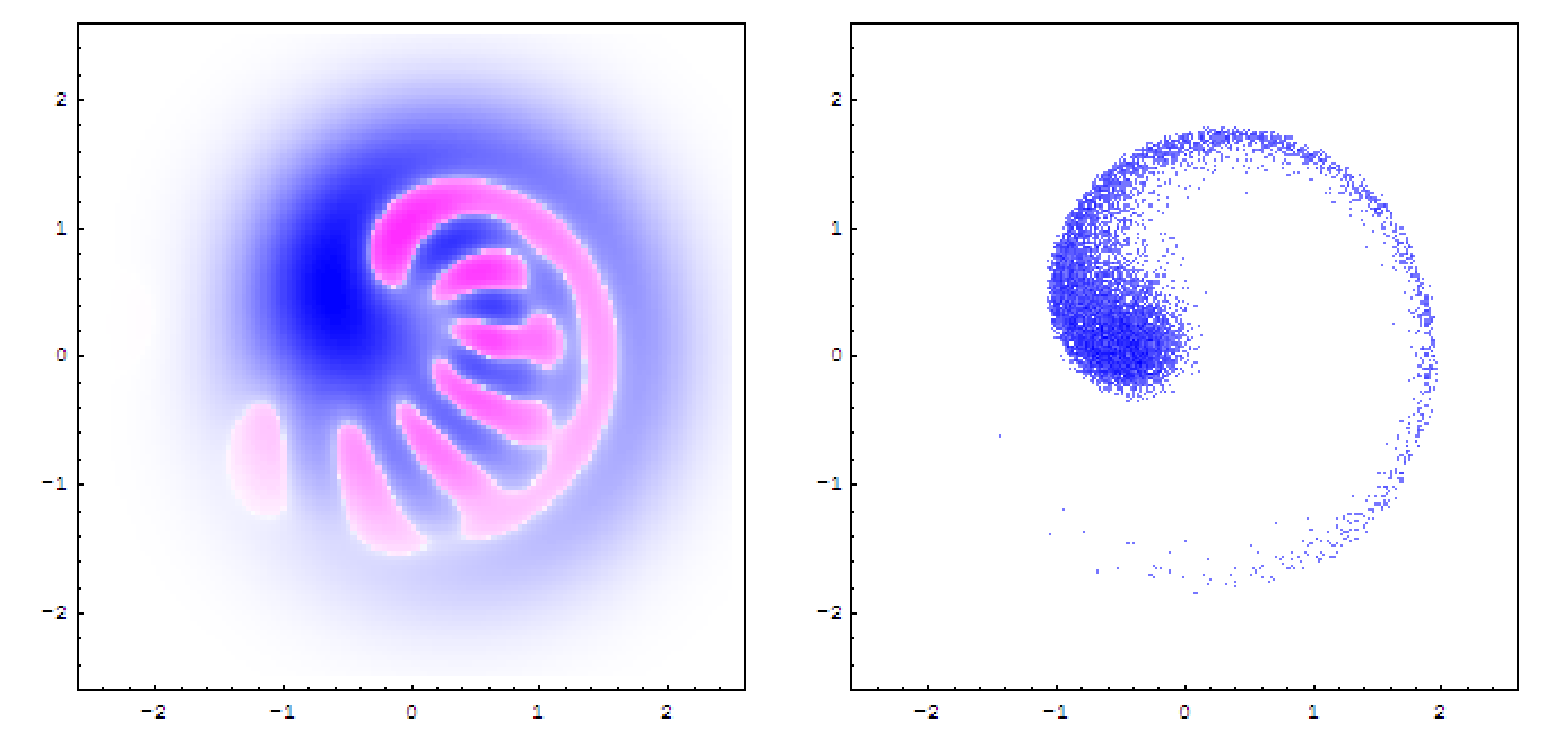}}
    \subfigure[\ $t=156T$]{
    \includegraphics[width=0.23\textwidth]{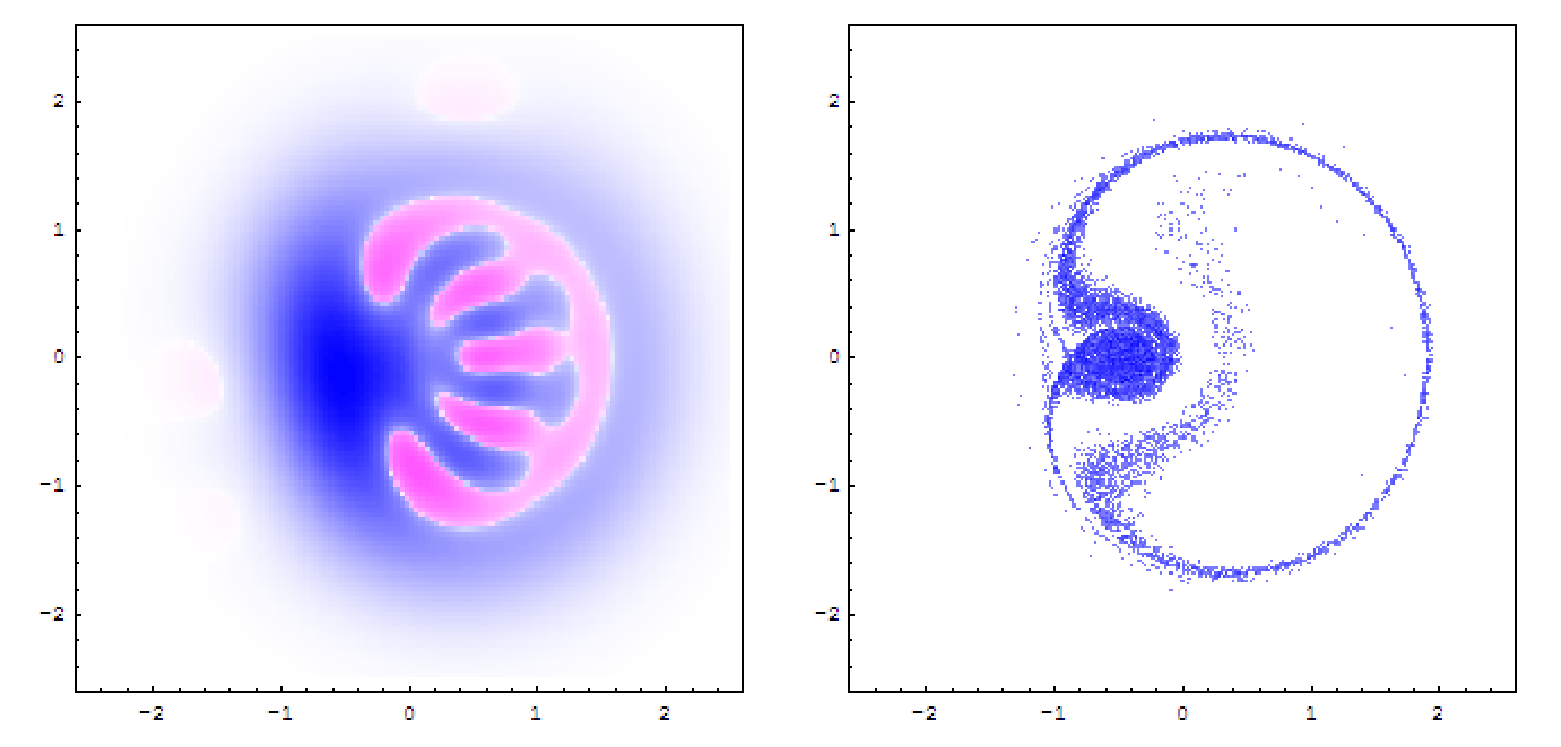}}
    \subfigure[\ $t=300T$]{
    \includegraphics[width=0.23\textwidth]{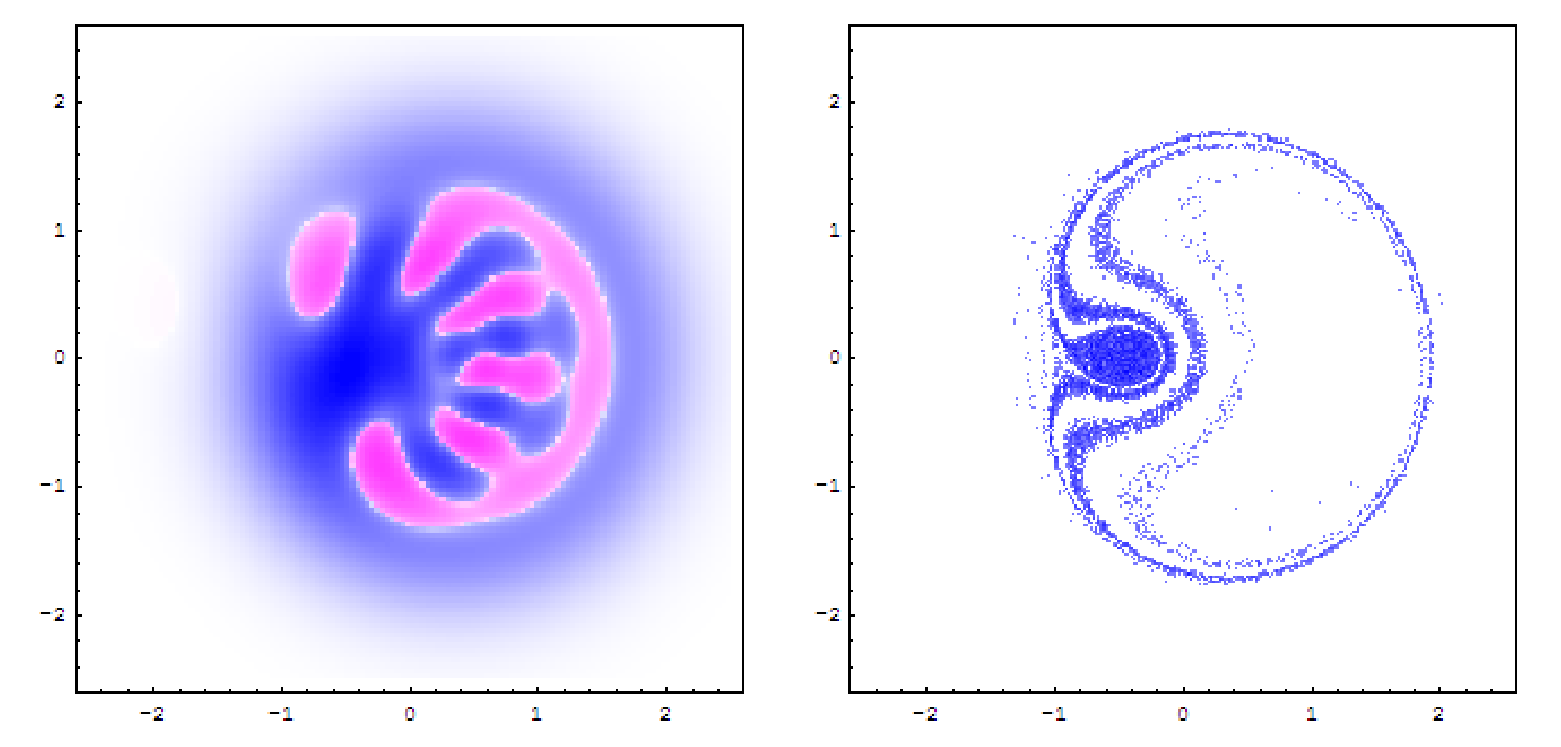}}
    \end{center}
    \caption{\label{nonlinear close 0.2} Isolated driven Duffing
      resonator with $\hbar=0.2$. Wigner functions (left) and
      classical distributions (right) of the initial minimal Gaussian
      wave packet and its evolution at three later times. A square of
      area $\hbar$ is shown at the bottom right corner of (a).}
\end{figure}

For the isolated quantum resonator we integrate the Schr\"{o}dinger
equation numerically by expanding the wave function in a truncated
harmonic-oscillator energy basis $|n\rangle$. From the wave function
we calculate the Wigner function~(\ref{wigner}), and compare it to the
integrated classical trajectories of the corresponding Gaussian
distribution of initial conditions.  Figure~\ref{nonlinear close 0.2}
shows the two distributions, calculated for $\hbar=0.2$, at $t=0$ and
at three later times, measured in units of driving period
$T=2\pi/\omega$. The Wigner functions are scaled to a power of $1/4$
to enhance weak features, where blue denotes positive values, and red
denotes negative values. A square of area $\hbar$ is given at the
bottom right corner of Fig.~\ref{nonlinear close 0.2}(a).

We clearly see similarities between the quantum and the classical
distributions, namely the positive backbone of the Wigner function
which resembles the classical distribution for short times, and the
maxima of the two distributions, which are roughly in the same
positions in phase space. Nevertheless, two differences prevent the
distributions from approaching each other even in the limit of
$\hbar\to 0$. These are the strong interference pattern in the Wigner
function and the infinitely fine structure that develops in the
classical distribution. As noted by Berry \cite{berry89}, the only way
to obtain a smooth transition from classical to quantum dynamics in
this case is to perform some averaging over a finite phase-space area
$\Delta x \Delta p \simeq \hbar$, taking into account the limited
precision of typical measurement devices.  Such averaging will smooth
out the delicate structure in the classical phase-space density and
will cancel out the interference fringes in the quantum Wigner
function, bringing the two into coincidence.
 
The need for  averaging is avoided once we consider the influence of
an environment. For the classical resonator we use a standard Langevin
approach, adding a velocity dependent dissipative term
$-\gamma\dot{x}$ and a time dependent random force $\delta F(t)$. The
latter is assumed to be a $\delta$-correlated Gaussian white noise,
related to the dissipative term according to the
fluctuation-dissipation theorem, thus defining a temperature
$T_{env}$. The quantum resonator is coupled linearly to a bath of
simple harmonic oscillators in thermal equilibrium at temperature
$T_{env}$.  This adds two terms to the quantum Hamiltonian
(\ref{Hamiltonian})---a Hamiltonian for the bath $H_{bath}$, and a standard
interaction Hamiltonian~\cite{santamore04}, taken to be of the
Caldeira-Leggett~\cite{caldeira_leggett83} type in the rotating-wave
approximation,
\begin{equation}
H_{int}=\sum_{i}(\kappa_{i}b_{i}a^{\dagger}+\kappa_{i}^{*}b_{i}^{\dagger}a),
\end{equation}
where the $\kappa_{i}$ are bilinear coupling constants, the $b_{i}$
are annihilation operators acting on the bath oscillators, and $a$
is the annihilation operator of the Duffing resonator. 

We consider the reduced density operator $\rho_{sys}$ of the Duffing
resonator by tracing over the bath degrees of freedom. By assuming the
interaction to be weak, and by employing the Markov approximation
which assumes the bath has no memory, we obtain a standard master
equation~\cite{santamore04,louisell73,gardiner_zoller04},
\begin{eqnarray}\label{master equation}\nonumber
  \dot\rho_{sys} = \frac{1}{i\hbar}[H_{sys},\rho_{sys}] -
  \frac{\gamma}{2}(1+\bar{n})(a^{\dagger}a\rho_{sys} 
+ \rho_{sys}a^{\dagger}a \\ 
  -2a\rho_{sys}a^{\dagger})-
  \frac{\gamma}{2}\bar{n}(aa^{\dagger}\rho_{sys} +
  \rho_{sys}aa^{\dagger}-2a^{\dagger}\rho_{sys}a), 
\end{eqnarray}
where $\bar{n} = (e^{\hbar\Omega/k_{B}T_{env}}-1)^{-1}$ is the
Bose-Einstein mean occupation, $\gamma = 2\pi g(\Omega)
|\kappa(\Omega)|^{2}$ is interpreted as the damping constant, and
$g(\Omega)$ and $\kappa(\Omega)$ are the density of states of the bath
and the coupling strength, respectively, both evaluated at the natural
frequency $\Omega=1$ of the resonator, because the nonlinearity is
small. Instead of integrating (\ref{master equation}) directly, we use
the \textit{Monte-Carlo wave function method}~\cite{MCWF,Plenio98},
which is more efficient computationally.

\begin{figure}[bt]
    \begin{center}
    \subfigure[\ Initial coherent state, $t=0$]{
    \includegraphics[width=0.23\textwidth]{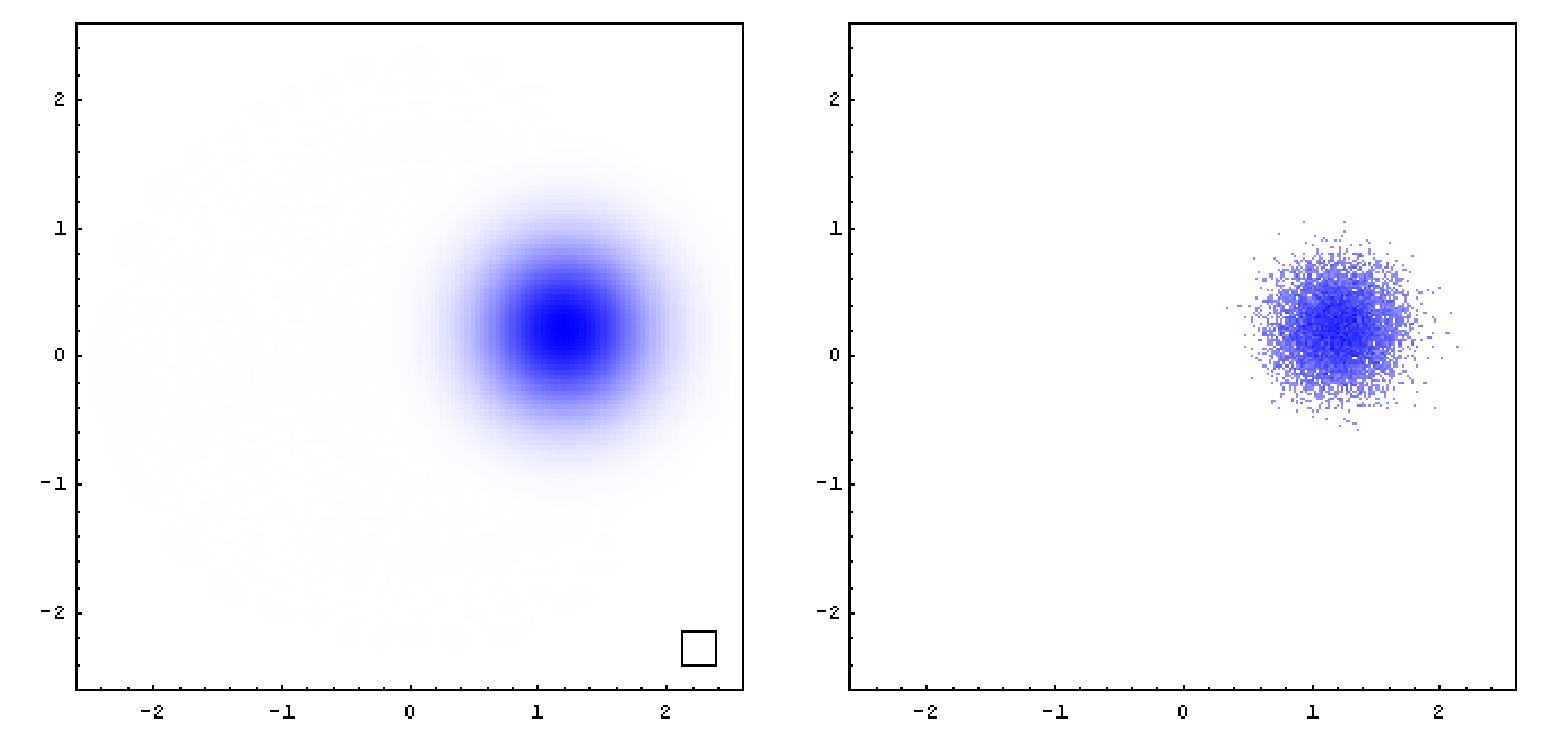}}
    \subfigure[\ $t=10T$]{
    \includegraphics[width=0.23\textwidth]{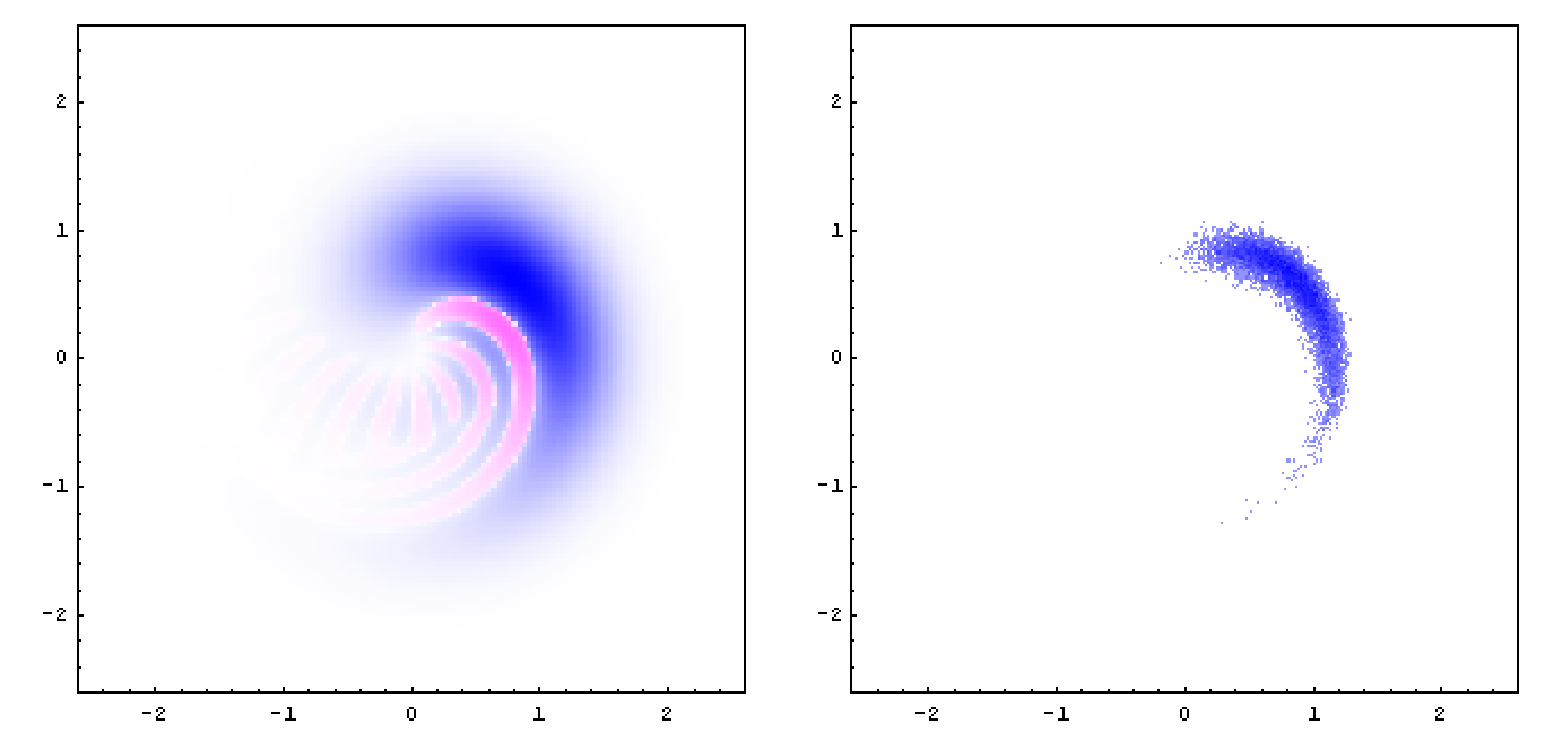}}
    \subfigure[\ $t=80T$]{
    \includegraphics[width=0.23\textwidth]{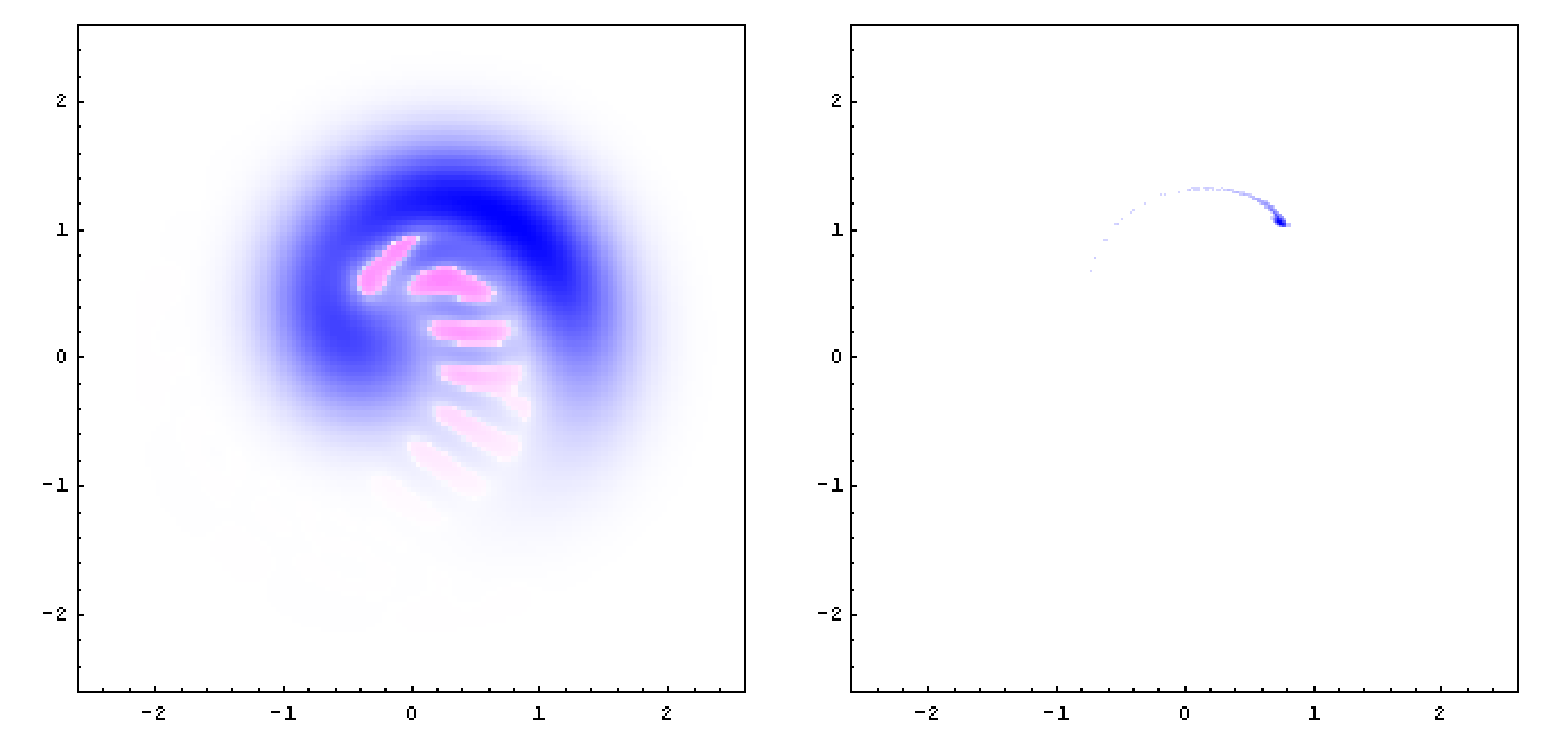}}
    \subfigure[\ $t=300T$]{
    \includegraphics[width=0.23\textwidth]{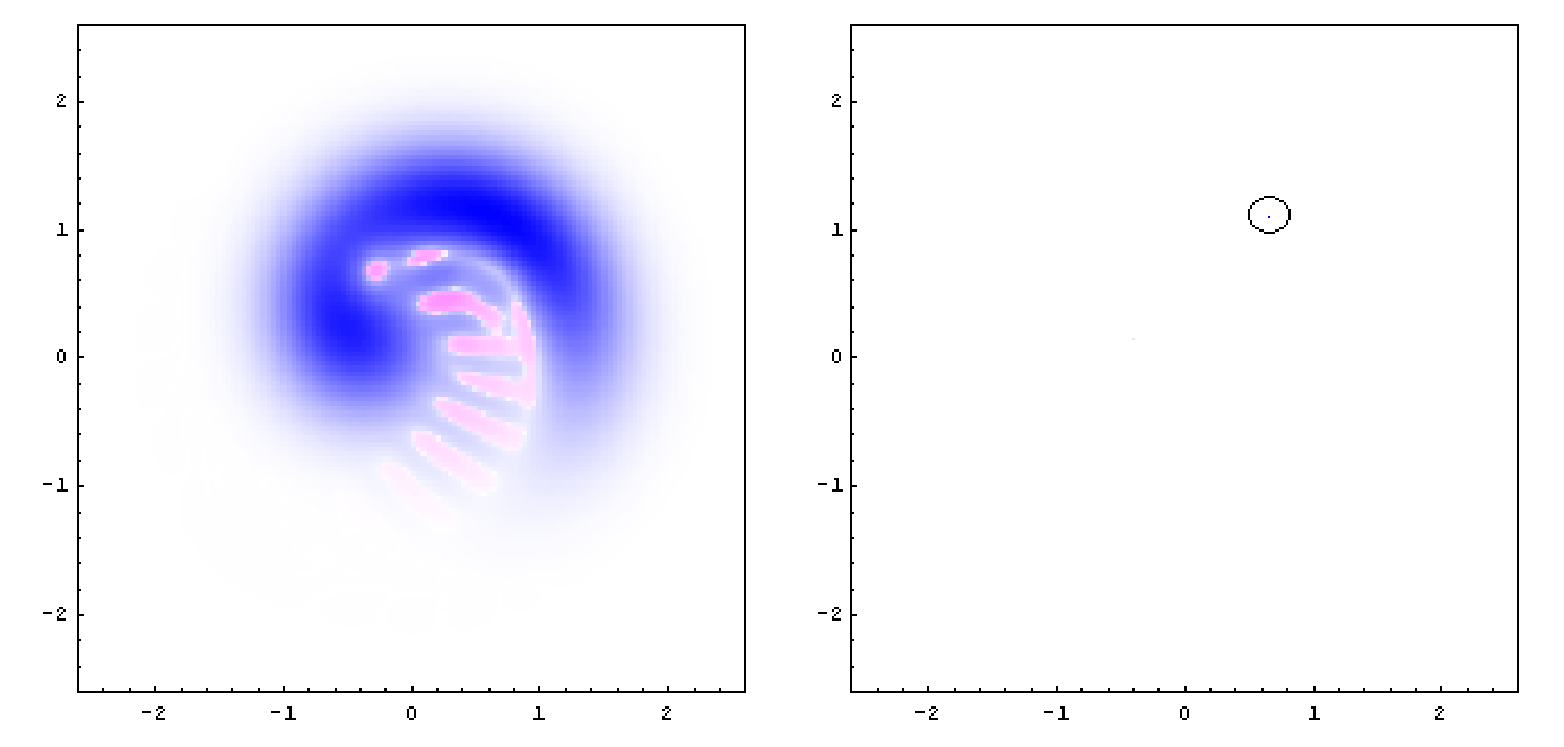}}
    \end{center}
    \caption{As in Fig.~\ref{nonlinear close 0.2}, but for a driven
      Duffing resonator with $\hbar=0.1$, coupled to a heat bath at
      $T_{env}=0$, with damping constant $\gamma=0.01$. The stable
      solution towards which all initial classical points flow is
      encircled in (d).}
    \label{nonlinear open kT0 hb0.1}
\end{figure}

Figure~\ref{nonlinear open kT0 hb0.1} shows the calculated results for
a Duffing resonator with $\hbar=0.1$, coupled to a heat bath at
temperature $T_{env}=0$. We take the initial state to be centered
around a point in phase space that under classical dissipative
dynamics flows towards the state of large amplitude oscillations. At
short times we see a general positive outline of the Wigner function
which is similar to the classical density, but it quickly deviates
from the classical distribution. The uncertainly principle prevents it
from shrinking to a point as in the classical dynamics. More
importantly, we clearly observe that the Wigner distribution has
substantial weight around the state of small-amplitude oscillations,
which is inaccessible classically for the chosen initial conditions at
$T_{env}=0$. The quantum resonator can switch between the two stable
states via tunneling or quantum activation~\cite{dykman06,dykman07},
although at this point it is impossible for us to distinguish between
these two processes.

\begin{figure}[bt]
    \begin{center}
    \subfigure[\ Initial coherent state, $t=0$]{
    \includegraphics[width=0.23\textwidth]{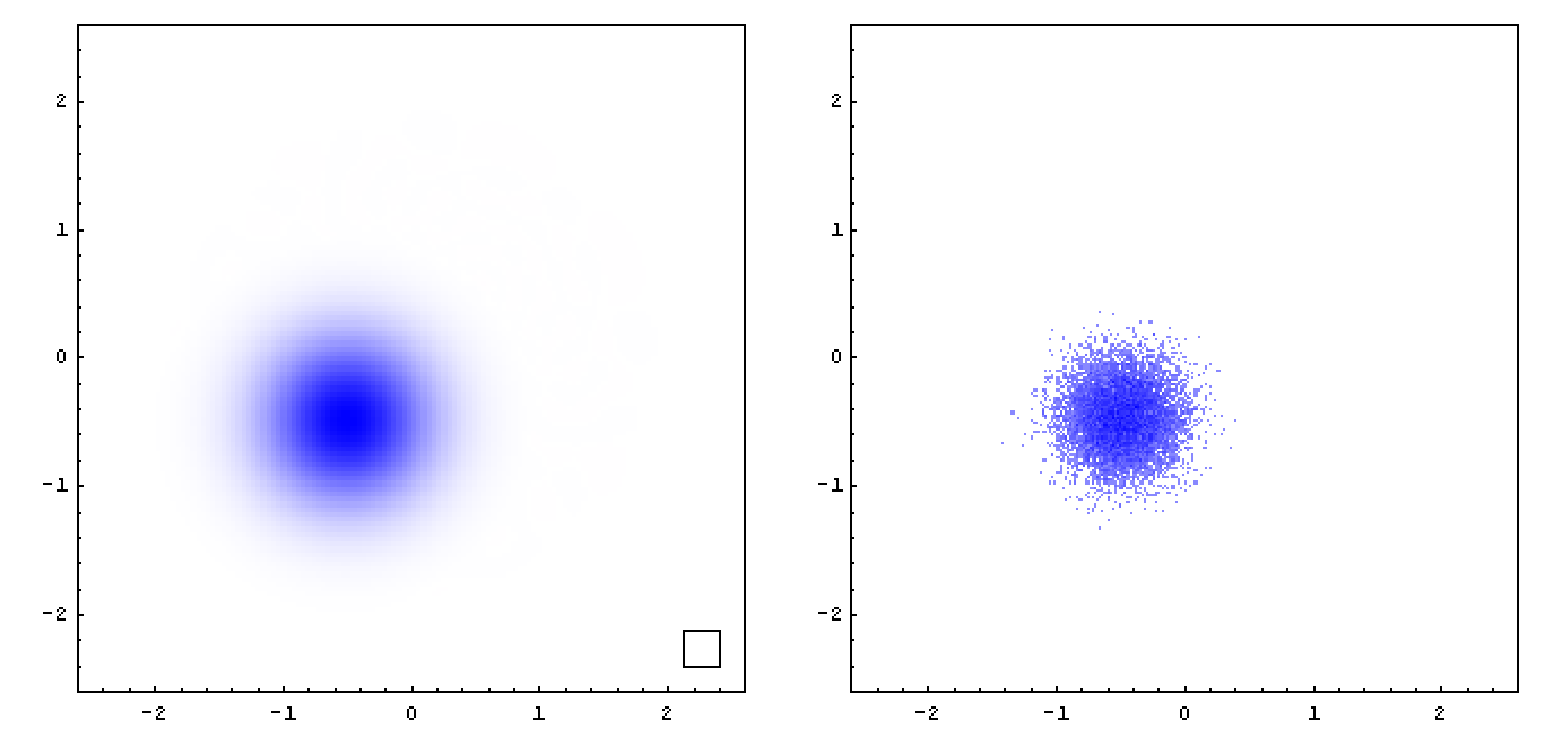}}
    \subfigure[\ $t=30T$]{
    \includegraphics[width=0.23\textwidth]{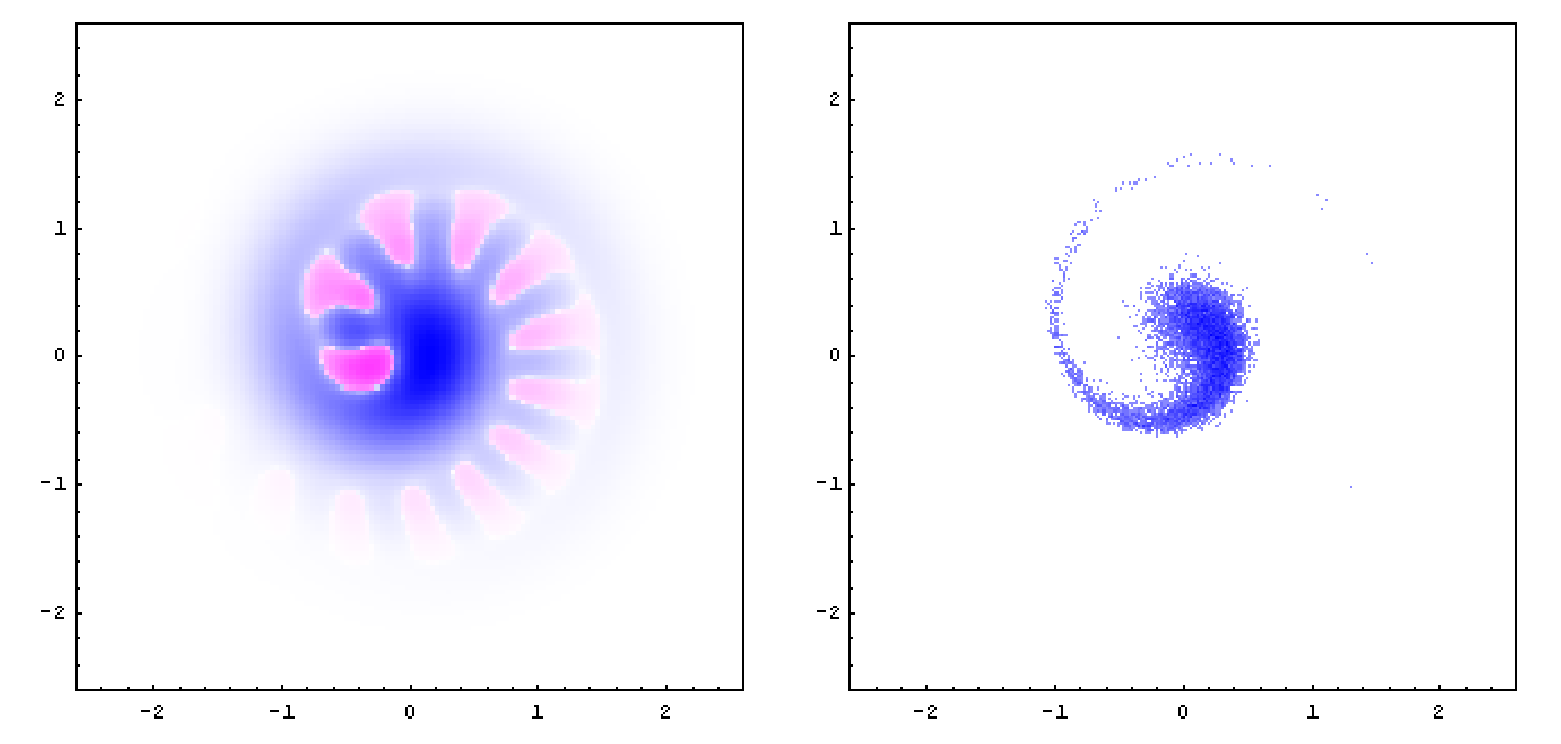}}
    \subfigure[\ $t=180T$]{
    \includegraphics[width=0.23\textwidth]{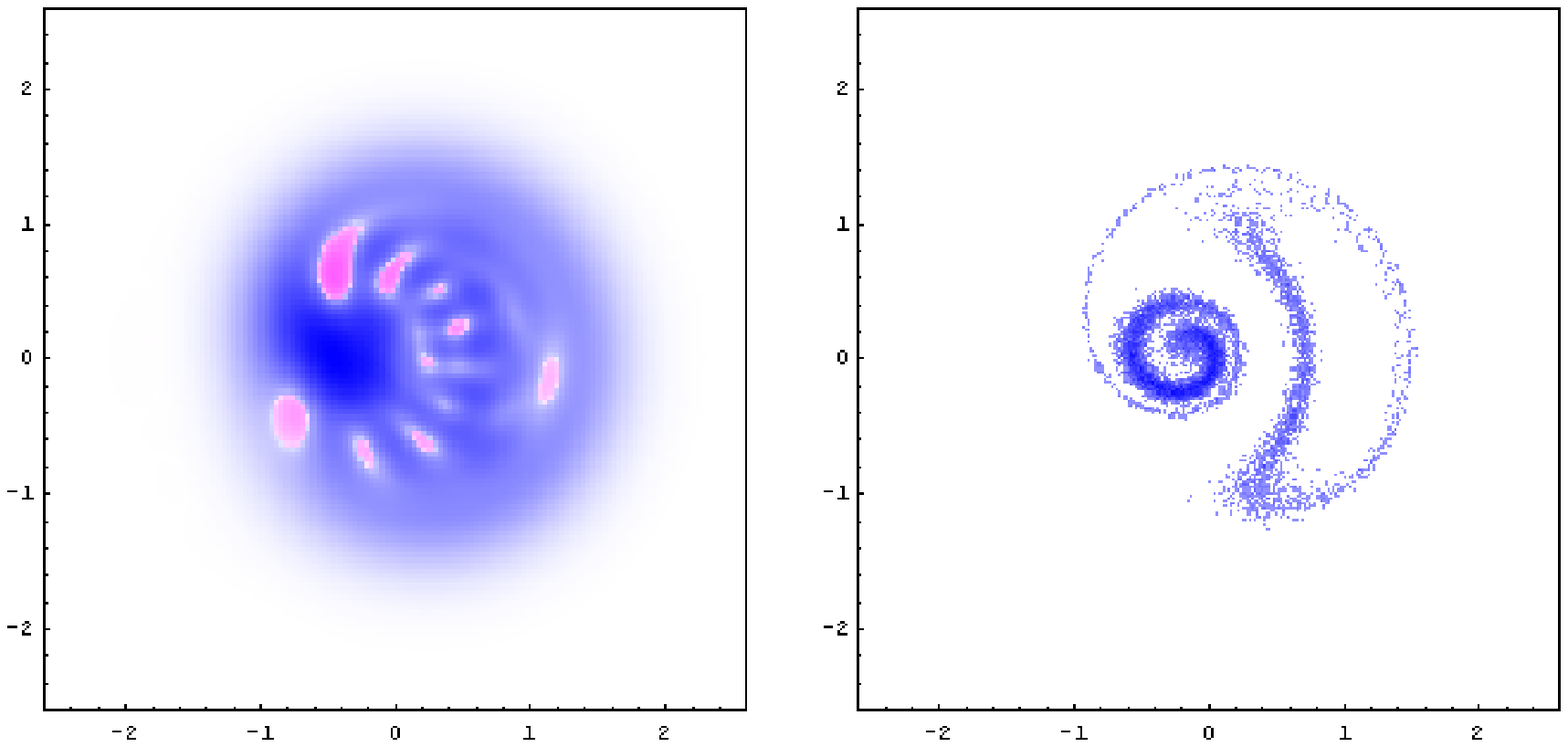}}
    \subfigure[\ $t=800T$]{
    \includegraphics[width=0.23\textwidth]{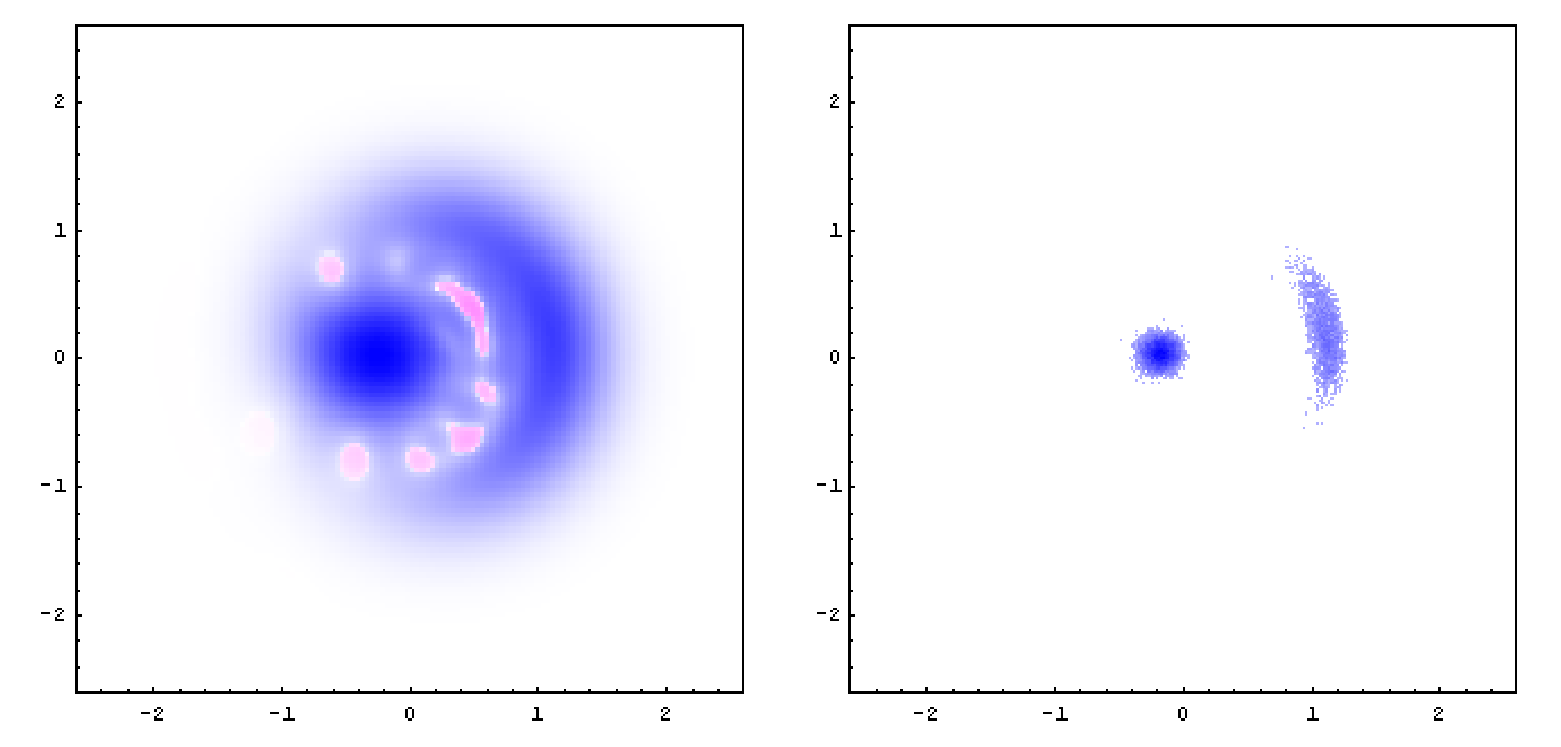}}
    \end{center}
    \caption{As in Fig.~\ref{nonlinear close 0.2}, but for a driven
      Duffing resonator with $\hbar=0.1$, coupled to a heat bath at
      $k_{B}T_{env}=2\hbar\Omega$, with damping constant
      $\gamma=0.001$.  The initial state is centered on the separatrix
      so that classically it splits towards both of the stable
      solutions.}
    \label{nonlinear open kT0.2 hb0.1}
\end{figure}

\begin{figure}[bt]
     \begin{center}
     \includegraphics[width=0.5\textwidth]{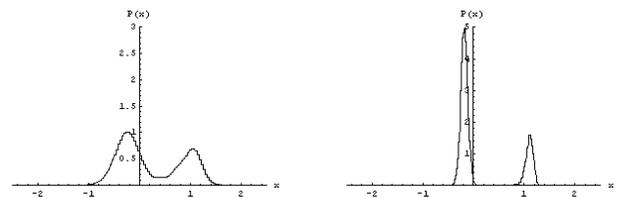}
     \end{center}
    \caption{Quantum (left)
      and classical (right) probability densities $P(x)$ for the
      phase-space distributions in Fig.~\ref{nonlinear open kT0.2
        hb0.1}(d).  The $y$ axes are scaled differently for better
      visualization.}
    \label{nonlinear open kT0.2 hb0.1PD}
\end{figure}

Figure~\ref{nonlinear open kT0.2 hb0.1} shows the calculated results
for a Duffing resonator with $\hbar=0.1$, coupled to a heat bath at a
finite temperature $k_{B}T_{env}=2\hbar\Omega$. This temperature is
obtained by adding a fluctuating force while reducing the dissipation,
thus requiring a longer time for the resonator to reach its final
steady state. Here we choose the initial state to straddle the
separatrix between regions in phase space that flow to the two stable
states. The interference pattern that develops for short times within
the general positive outline in the Wigner function is soon destroyed
by decoherence, and becomes erratic in space and time. At long times,
both distributions peak around the two stable states, nevertheless
they differ significantly. The classical density is tightly localized
around the two solutions with no overlap, indicating that $T_{env}$ is
too small to induce thermal switching between the states, as was
observed in a recent experiment~\cite{aldridge05}. The Wigner
function, on the other hand, is spread out in phase space, indicating
that $\hbar$ is sufficiently large for the quantum resonator to switch
between the two states via tunneling or quantum
activation~\cite{dykman06,dykman07}. This is demonstrated more clearly
in Fig.~\ref{nonlinear open kT0.2 hb0.1PD}, which shows the
probability distributions $P(x)$ at the steady state. We find that
only for temperatures as high as $k_{B}T=17\hbar\Omega$ does the
classical phase-space distribution become as wide as the Wigner
function is at $k_{B}T=2\hbar\Omega$, as shown in Fig.~\ref{effective
  temperature}. Thus, in a real experiment, evidence for
quantum-mechanical dynamics can be demonstrated as long as temperature
and other sources of noise can be controlled to better than an order
of magnitude.

\begin{figure}[bt]
    \begin{center}
    \subfigure[\ Phase space densities, $t=800T$]{
    \includegraphics[width=0.4\textwidth]{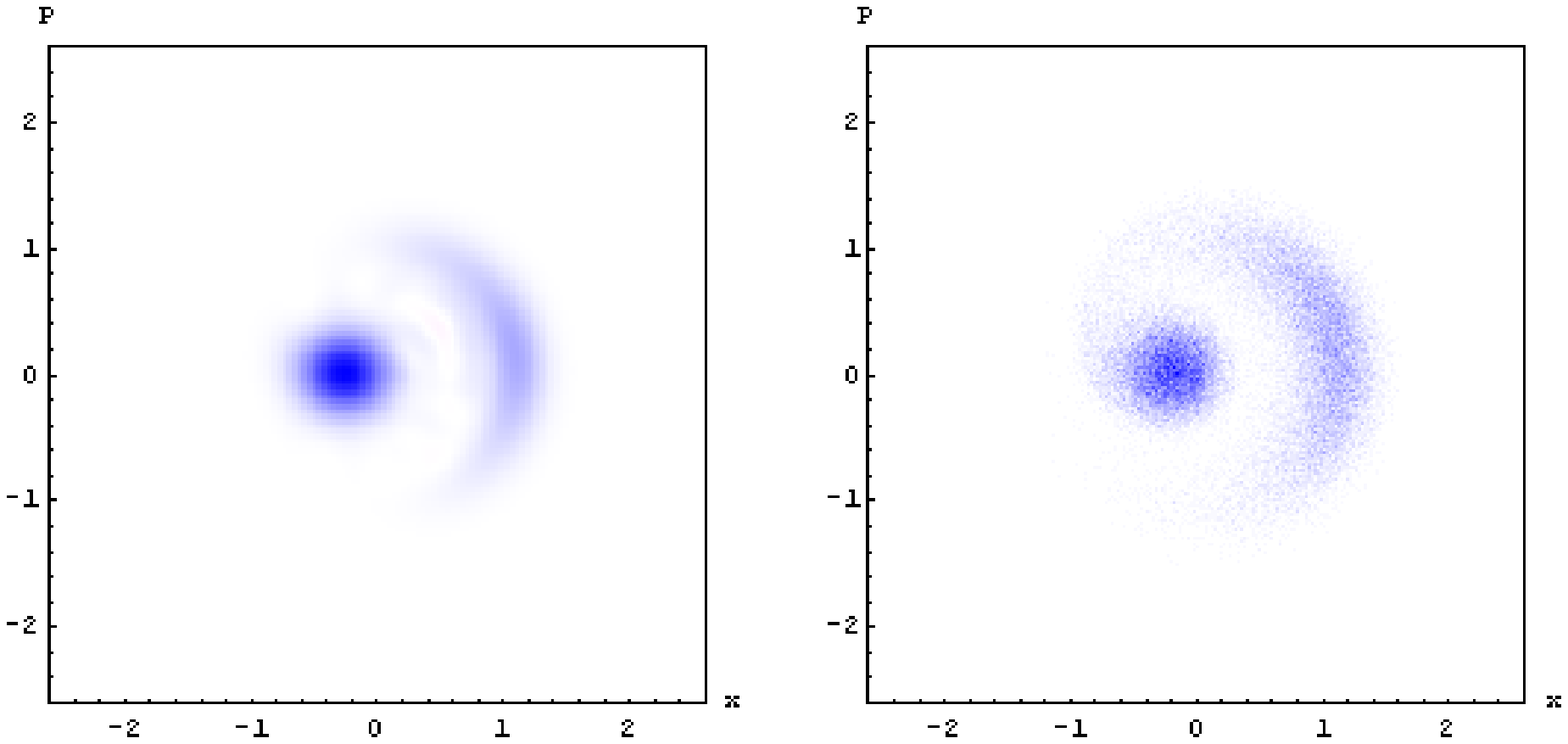}}
    \subfigure[\ Probability densities $P(x)$ at $t=800T$]{
    \includegraphics[width=0.4\textwidth]{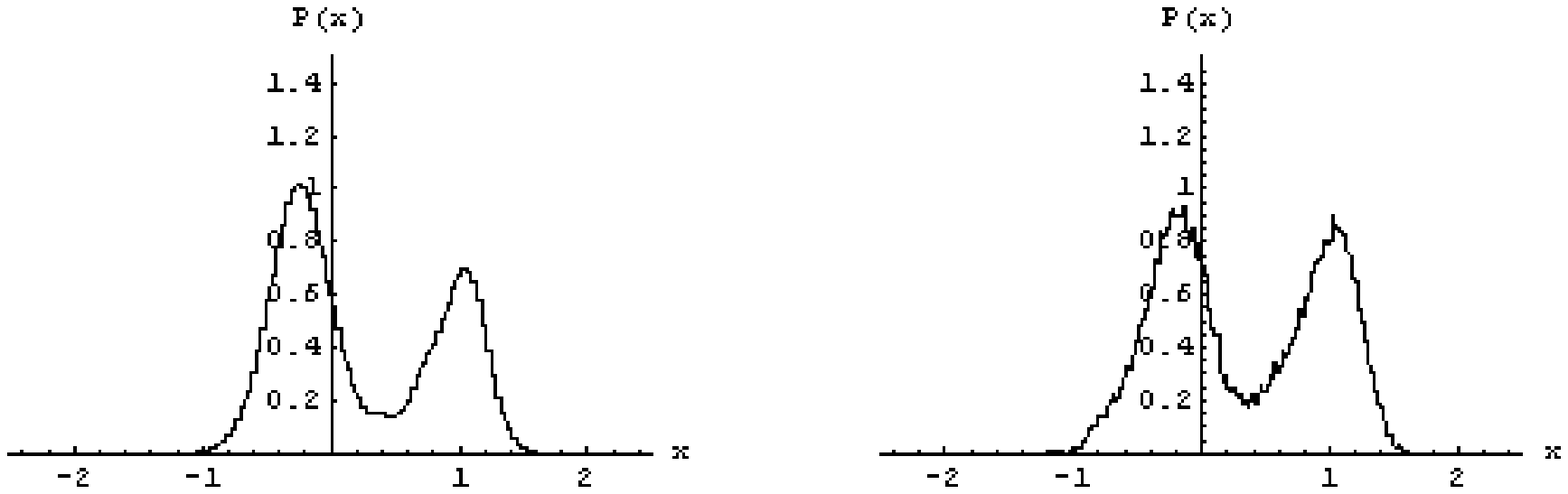}}
    \end{center}
    \caption{As in Figs.~\ref{nonlinear open kT0.2 hb0.1}(d) and
      \ref{nonlinear open kT0.2 hb0.1PD}, only that the classical
      calculation (right) is for $k_{B}T_{env}=17\hbar\Omega$,
      yielding similar phase-space distributions (a) and probability
      densities (b). Both distributions in (a) are left unscaled.}
    \label{effective temperature}
\end{figure}

We have found the evidence we were seeking for a mechanical system
entering the quantum domain with effective $\hbar=0.1$ and
$k_{B}T_{env}=2\hbar\Omega$. It stems from the ability of a quantum
Duffing resonator to switch between the two stable states, having
finite probability of being in between the two states, while the
classical resonator cannot. We intend to study individual quantum
trajectories of the resonator to determine whether switching takes
place via tunneling or quantum activation.

An effective $\hbar\simeq 0.1$ implies that the area in phase space where
the dynamics takes place, is on the order of $10\hbar$.  This area is
roughly $x_{max}p_{max} \simeq m\Omega {a_c}^2$, where $a_c$---the
critical oscillation amplitude for the onset of bistability---is
proportional to $d/\sqrt{Q}$~\cite{Inna05}, with $d$ being the diameter
of the resonator, and $Q$ its quality factor. In current NEMS
resonators~\cite{Inna05,Inna06} $m\simeq 10^{-18}$kg, $\Omega\simeq
10^{8}$Hz, and $a_{c}\simeq 10^{-9}$m, yielding an effective $\hbar$
of $10^{-6}$, which is well within the classical domain. Yet, if we
consider suspended nanotubes~\cite{yaish04,vanderzant06}, then
optimistic values may give $m\simeq 10^{-21}$kg, $\Omega\simeq 10^{8}$Hz and
$a_{c}\simeq 10^{-11}$m, for which the effective $\hbar\simeq 10$,
which is much better than needed, while requiring a temperature of
$T_{env}\simeq10$mK. We therefore believe that it should be possible
to realize our calculations in a real experiment in the near future.

The authors thank Michael Cross, Mark Dykman, Jens Eisert, Victor
Fleurov, Inna Kozinsky, and Keith Schwab for illuminating discussions.
RS is grateful to the Lion Foundation for supporting his stay at Tel
Aviv University as an exchange student.  This research is supported by
the U.S.-Israel Binational Science Foundation under Grant No.~2004339,
and by the Israeli Ministry of Science.

\bibliography{itamar}

\end{document}